\documentclass[11pt]{article}
\usepackage{moriond,epsfig}

\usepackage{amsmath,amssymb}

\def\be{\begin{equation}}
\def\ee{\end{equation}}
\def\bea{\begin{eqnarray}}
\def\eea{\end{eqnarray}}

\begin{document}
\hfill {\tt CERN-PH-TH/2012-121}

\vspace*{4cm}
\title{SUSY and a 125 GeV Scalar}

\author{ F. Mahmoudi }

\address{CERN Theory Division, Physics Department, CH-1211 Geneva 23, Switzerland}

\address{Clermont Universit{\'e}, Universit\'e Blaise Pascal, CNRS/IN2P3,\\
LPC, BP 10448, 63000 Clermont-Ferrand, France}

\maketitle\abstracts{
An excess of events at a mass of $\sim$ 125 GeV has been reported by both ATLAS and CMS collaborations using 5~fb$^{-1}$ of data. If this excess of events is confirmed by further searches with more data, it will have extremely important consequences in the context of supersymmetric extensions of the Standard Model. We show that for a Standard Model like Higgs boson with a mass 122.5 $< M_h <$ 127.5 GeV, several unconstrained or constrained MSSM scenarios would be excluded, while the parameters of some other scenarios would be severely restricted.
}
%
\section{Introduction}
The search for supersymmetry (SUSY) constitutes the main focus of new physics searches at the LHC. With no signal detected so far, these searches have imposed strong limits mainly in the constrained 
SUSY scenarios. However a lot of solutions compatible with all present limits still remain  \cite{Arbey:2011un,Sekmen:2011cz}, and even if the bounds are becoming stronger, it is not possible to validate or falsify supersymmetry as a viable extension of the Standard Model (SM). An alternative path to tightly constrain and test the MSSM at the LHC is through the Higgs sector. Searches for the Higgs boson have now narrowed down the possible window for an SM--like Higgs to only a few GeV around 125 GeV, where an excess over background in several channels has been observed \cite{ATLAS:2012ae,Chatrchyan:2012tx}. If this excess is confirmed, it will have extremely important consequences for the MSSM, and we highlight here some of the main outcomes. This write-up is based on \cite{Arbey:2011ab,Arbey:2011aa,abdm} where more details about the analyses can be found.
%
\section{Higgs mass predictions}
In the SM, the Higgs boson mass is basically a free parameter. In the MSSM however the lightest Higgs mass is bounded from above:
\begin{equation}
M^{max}_h \approx M_Z |\cos 2 \beta| + \mbox{radiative corrections} \lesssim 110-135\mbox{ GeV .}
\end{equation}
Imposing $M_h$ places therefore very strong constraints on the MSSM parameters through their contributions to the radiative corrections. The most sensitive parameters to the Higgs mass are $\tan\beta$, the CP--odd Higgs mass $M_A$, the SUSY breaking parameter $M_S = \sqrt{m_{\tilde t_1}m_{\tilde t_2}}$ and the mixing parameter in the stop sector, $X_t= A_t -\mu/\tan\beta$. In the decoupling regime, the Higgs mass can be approximated by:
\begin{equation}
M_{h}^2 \overset{M_A \gg M_Z} \approx M_Z^2 \cos^2 2\beta + \dfrac{3 m_t^4}{2 \pi^2 v^2} \left[ \log \dfrac{M_S^2}{m_t^2} + \dfrac{X_t^2}{M_S^2}\left( 1 - \dfrac{X_t^2}{12 M_S^2} \right)\right]\;.
\end{equation}
The maximal Higgs mass can therefore be reached in the decoupling regime with large $M_A$, for large values of $\tan\beta$ ($\gtrsim 10$), for heavy stops (large $M_S$) and for the stop mixing parameter $X_t=\sqrt{6}M_S$ corresponding to the so--called maximal mixing scenario.
%
\section{Implications for MSSM}
\begin{figure}[!t]
\centering
\includegraphics[width=10.cm]{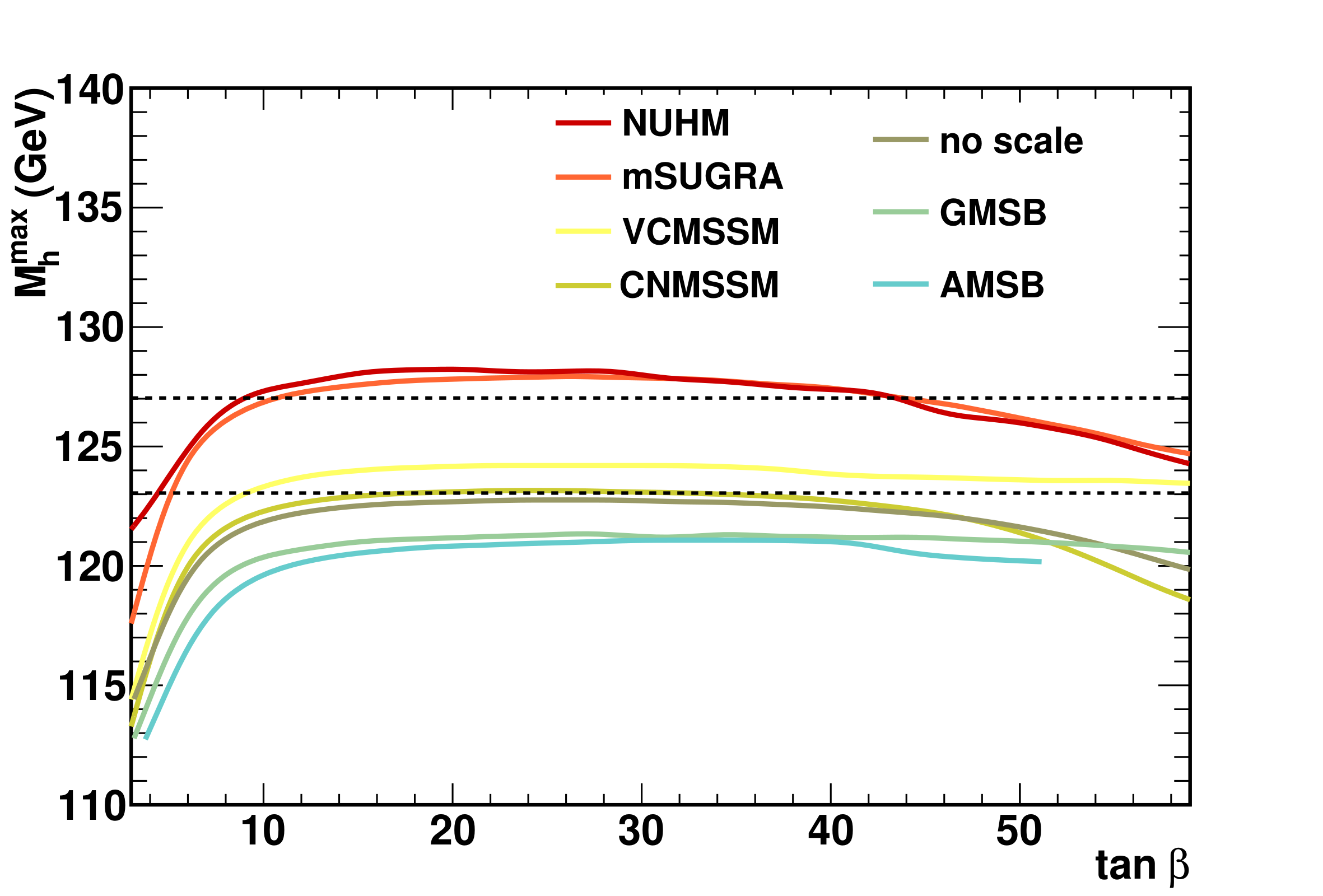}
\caption{The maximal $M_h$ defined as the value for which 99\% of the scan points have a mass smaller than it, shown as a function of $\tan\beta$ for various constrained MSSM models.}
  \label{fig:mhmax}
\end{figure}

\begin{figure}[!t]
\centering
\includegraphics[width=7.5cm]{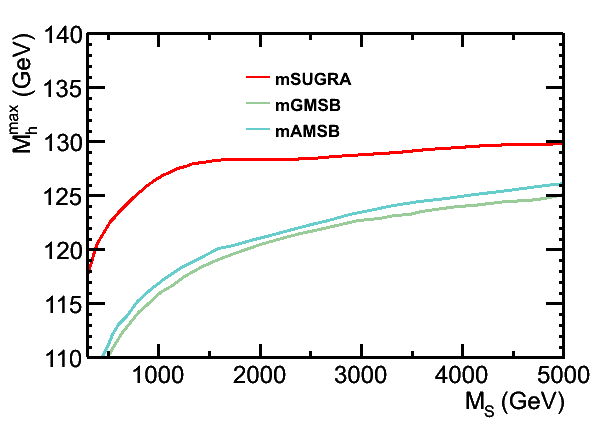}
\caption{The maximal $M_h$ in function of $M_S$ in mAMSB and mGMSB. For comparison the results in mSUGRA are also given.}
  \label{fig:mhms}
\end{figure}

We first consider the implications of a Higgs boson mass determination for the constrained MSSM scenarios. For this purpose, we perform extensive scans in the parameter spaces of several constrained MSSM scenarios and generate the spectra using {\tt Suspect} \cite{suspect}. Since the various parameters which enter the radiative corrections to the MSSM Higgs sector are not all independent, it is not possible to freely tune the relevant weak--scale parameters to obtain a given value of $M_h$. In particular, a Higgs mass of around 125 GeV can have drastic consequences on the constrained MSSM scenarios as can be seen from Fig. \ref{fig:mhmax} where we restricted $M_S$ to be below 3 TeV in order to have an ``acceptable'' finetuning. Table \ref{tab:mhmax} gives the maximal Higgs mass reached in different constrained MSSM scenarios. 
Only scenarios such as mSUGRA and NUHM can provide a sufficiently large mass for the Higgs. On the other hand, scenarios such as AMSB and GMSB are disfavoured in their minimal versions as the term $A_t/M_S$ is rather small and we are almost in the no--mixing regime. However, by relaxing the condition on $M_S$, larger values for $M_h$ could also become possible in these scenarios (at the cost of increasing the finetuning) as can be seen in Fig.~\ref{fig:mhms}. The definition of the different models and more detailed discussions can be found in \cite{Arbey:2011ab}.
\begin{table}[!t]
\begin{center}
\begin{tabular}{|l|c|c|c||c|c|c|c|}\hline
model & AMSB & GMSB & mSUGRA & no-scale & CNMSSM & VCMSSM& NUHM \\ \hline
$M_h^{\rm max}$ & 121.0 & 121.5 & 128.0 & 123.0 & 123.5 & 124.5 & 128.5 \\ \hline
\end{tabular}
\end{center}
\caption{Maximal $M_h$ value (in GeV) in various constrained MSSM scenarios.}
  \label{tab:mhmax}
\end{table}

\begin{figure}[!t]
\centering
\includegraphics[width=7.5cm]{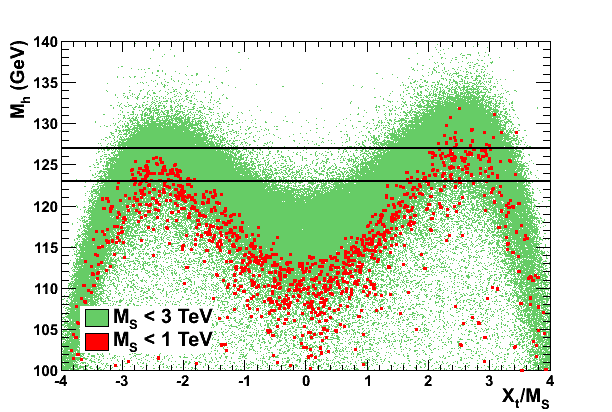}\includegraphics[width=7.5cm]{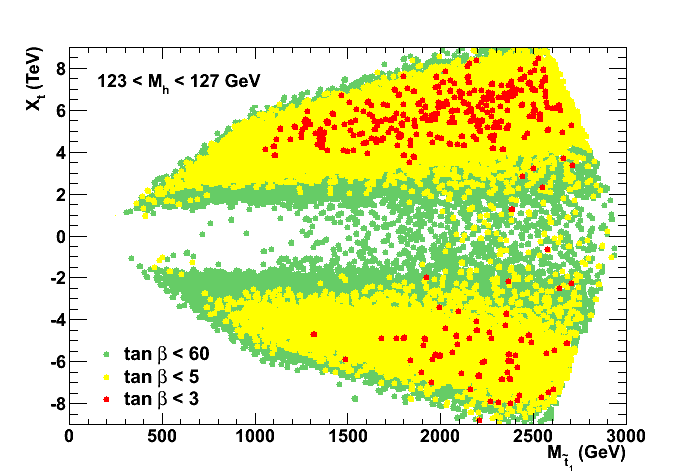}
\caption{Light Higgs mass $M_h$ as a function of $X_t/M_S$ in the pMSSM (left) and the contours for 123 $< M_h <$ 127~GeV in the $(M_{\tilde t_1},X_t)$ plane for some
selected ranges of $\tan\beta$ values (right).}
  \label{fig:pmssm}
\end{figure}

Next we study the consequences for the phenomenological MSSM (pMSSM) \cite{pmssm}, where no universal boundary condition is assumed. The pMSSM with CP and R-parity conservation involves 19 free parameters. To study the pMSSM, we perform flat scans over the parameters as described in \cite{Arbey:2011ab}. 
In Fig.~\ref{fig:pmssm} we show the light Higgs mass as a function of $X_t/M_S$.
As can be seen from the figure, scenarios with large $X_t/M_S$ values and, in particular, those close to the maximal mixing $A_t/M_S \approx \sqrt 6$ are favoured by the Higgs mass constraints. On the other hand, the no--mixing scenario ($X_t \approx 0$) is strongly disfavoured for $M_S \lesssim 3$ TeV, and the typical mixing scenario needs large $M_S$ and moderate to large $\tan\beta$ values. 

The requirement of having the lightest Higgs in the range 123 $<M_h<$ 127 GeV does not necessarily correspond to very heavy stop masses. Indeed, stops as light as 350 GeV can be possible in the pMSSM as can be seen from the right hand side of Fig.~\ref{fig:pmssm}.
\begin{figure}[!t]
\centering
\includegraphics[width=6.5cm]{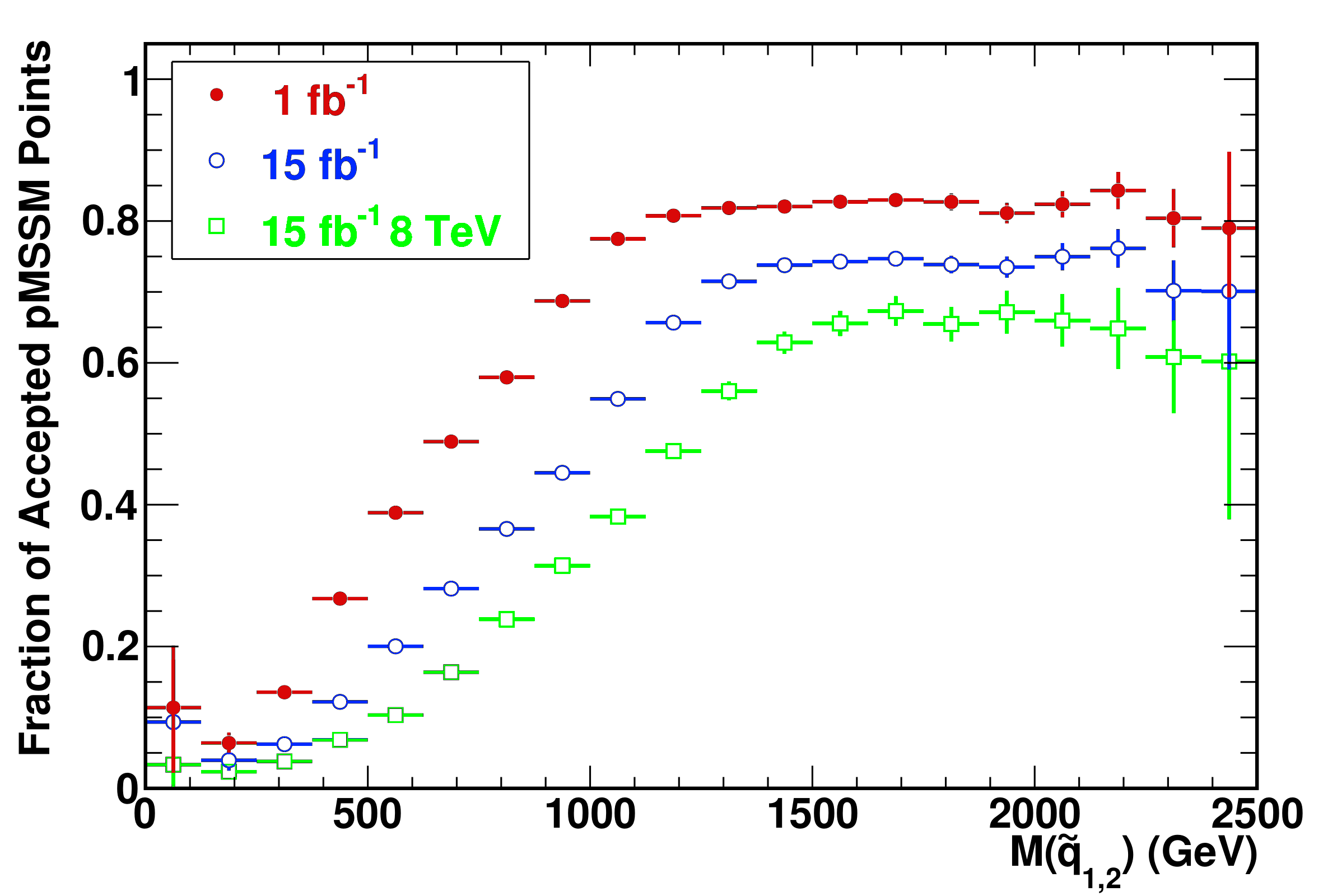}
\includegraphics[width=6.5cm]{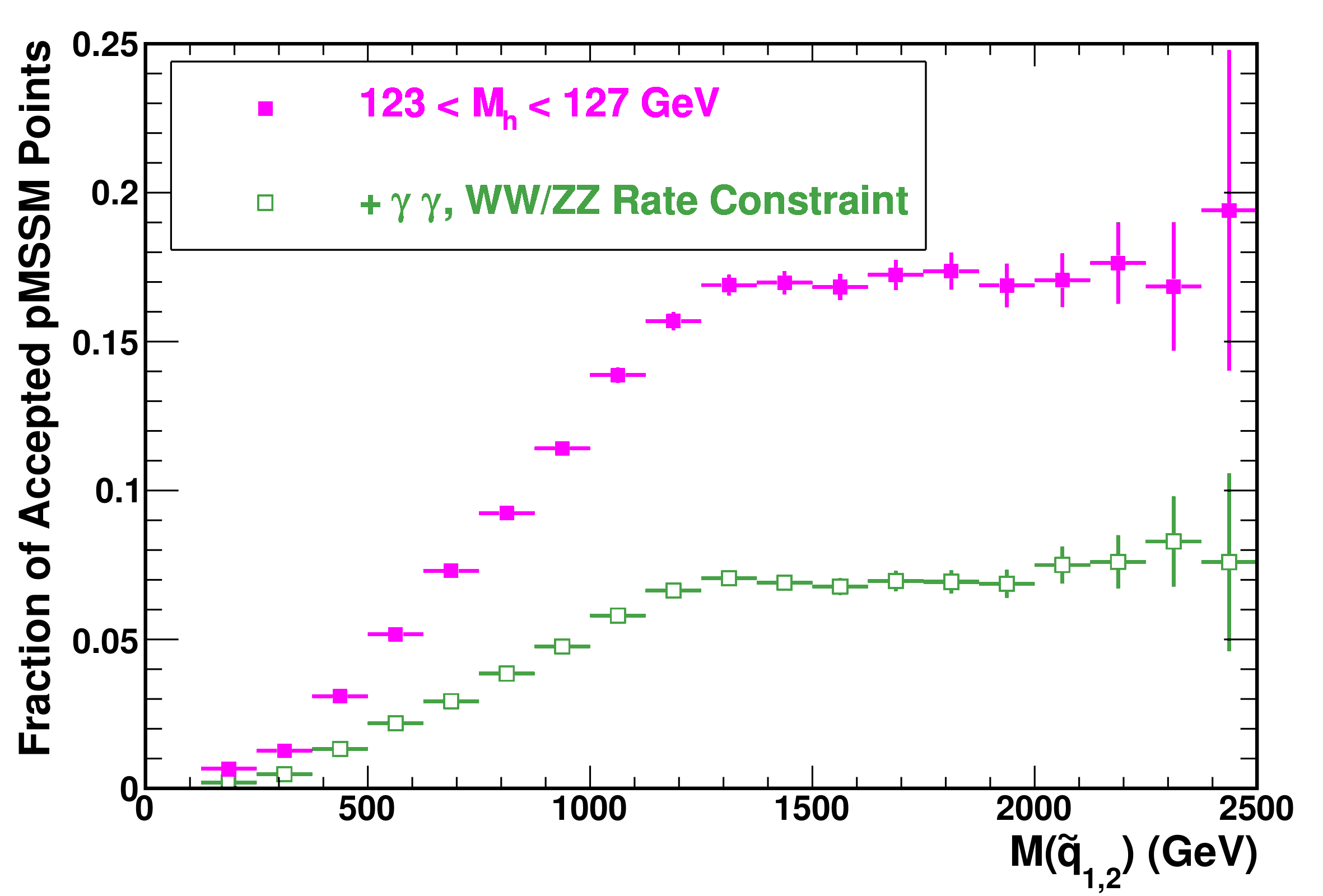}
\includegraphics[width=6.5cm]{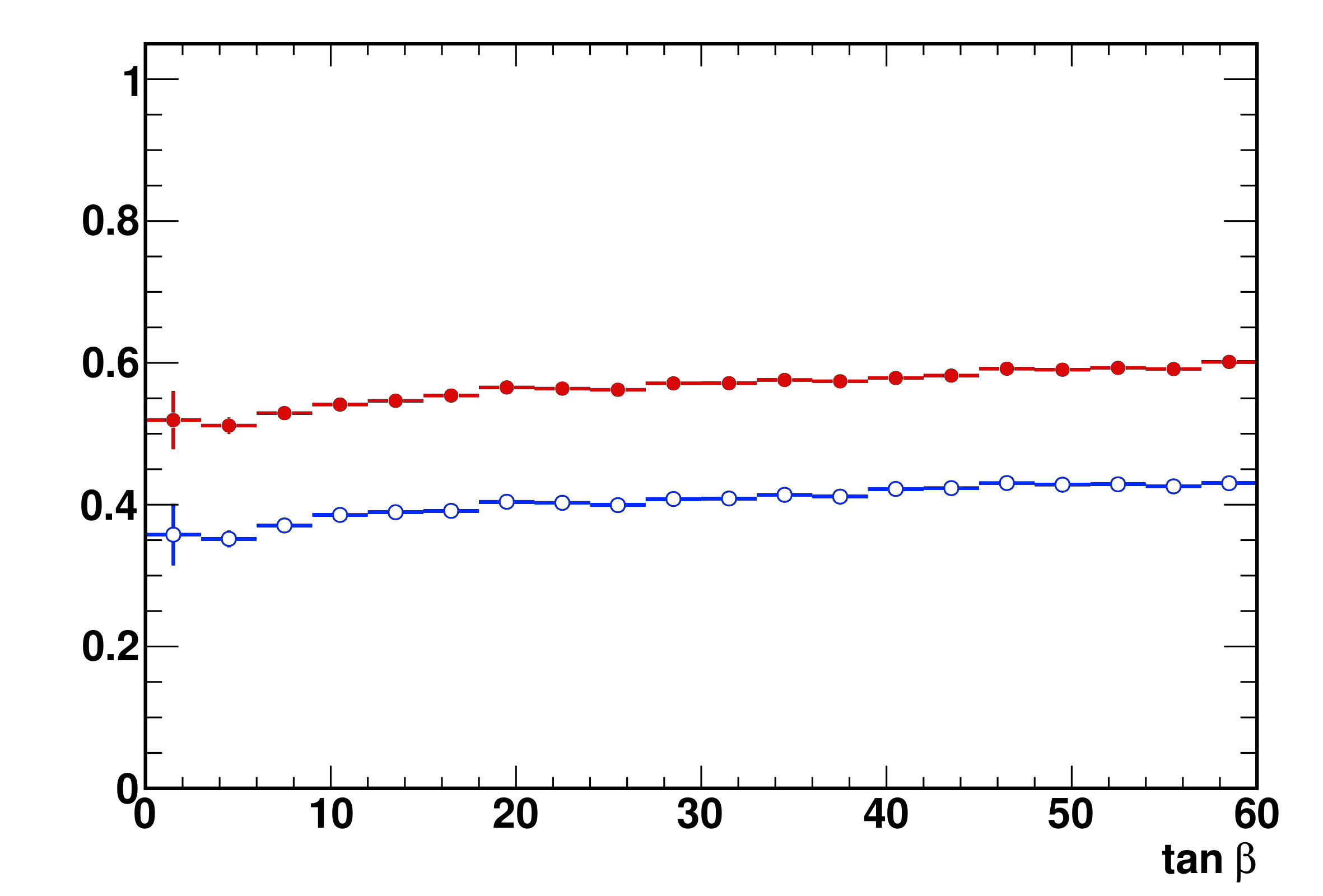}
\includegraphics[width=6.5cm]{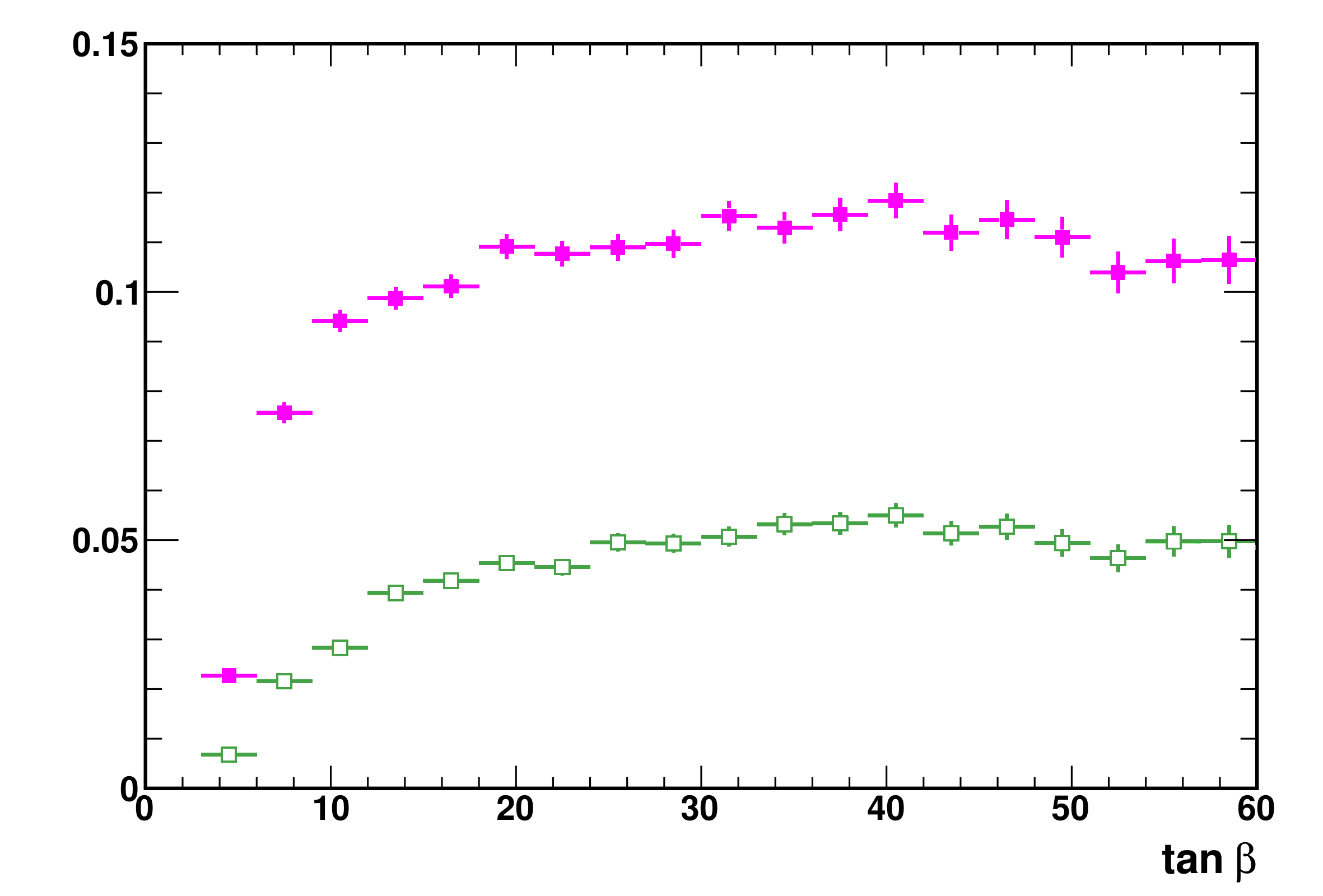}
\caption{Fraction of accepted pMSSM points not excluded by the SUSY searches on 1 and 15~fb$^{-1}$ 
of LHC data as a function of the mass of the lightest squark of the first two generations 
(upper left) and of $\tan \beta$ (lower left), and with the Higgs constraint in addition in the right hand side.}
  \label{fig:pmssm2}
\end{figure}

It is possible to further constrain the parameter space by combining the information from direct Higgs and SUSY searches with those from flavour physics and dark matter. We calculate the flavour constraints, dark matter relic density and muon ($g-2$) using {\tt SuperIso Relic} \cite{superiso_relic}, the SUSY spectra using {\tt Softsusy} \cite{softsusy}, the Higgs decay rates using {\tt HDECAY} \cite{hdecay}. To assess the observability of each of the pMSSM scan points we generate events with {\tt PYTHIA} \cite{pythia} which are passed through fast detector simulation using {\tt Delphes} \cite{delphes}.
More details about the scans and the employed tools can be found in \cite{Arbey:2011un,Arbey:2011aa}.

Fig.~\ref{fig:pmssm2} shows the fraction of pMSSM points which pass all the constraints and are compatible with the direct SUSY searches by CMS with 1~fb$^{-1}$ of data \cite{Chatrchyan:2011zy} and the projection for 15~fb$^{-1}$ as a function of the masses of the lightest squarks of the first two generations $\tilde q_{1,2}$ and $\tan\beta$ \cite{Arbey:2011un,Arbey:2011aa}. As can be seen from the plots 15~fb$^{-1}$ of LHC data should provide a powerful constraint to MSSM solutions. In the right hand side of Fig.~\ref{fig:pmssm2}, we impose in addition constraints from the discovery of an SM--like Higgs at the LHC. A comparison between the right and left handed plots shows that the fraction of accepted points is strongly reduced, and values of $\tan\beta \leq 6$ become disfavoured, while the shape of the mass distributions of squarks is not significantly affected.

\begin{figure}
\centering
\includegraphics[width=7.5cm]{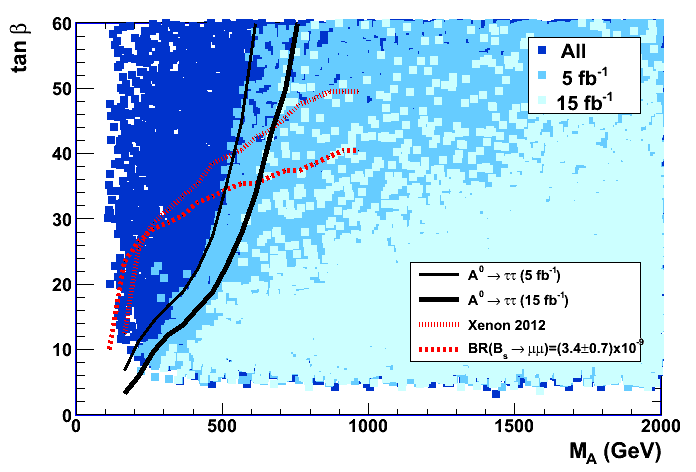}
\includegraphics[width=7.5cm]{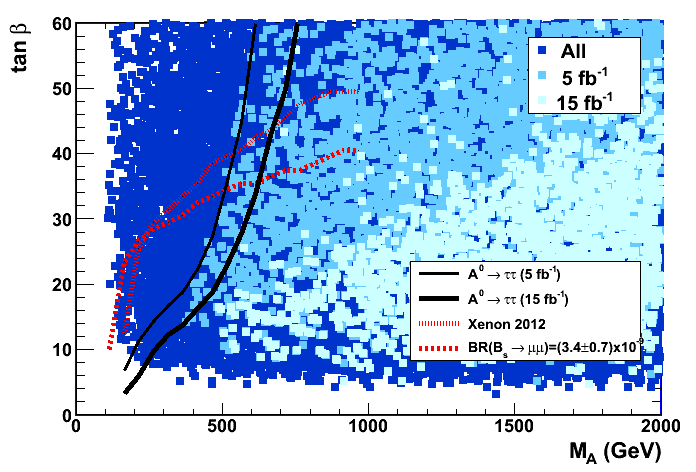}
\caption{pMSSM points in the ($M_A$, $\tan \beta$) plane giving 123 $< M_H <$ 127~GeV. The different shades of blue show the points in the pMSSM without cuts and those allowed by the 2011 data and by the projected 2012 data, assuming no signal beyond the lightest Higgs boson is observed. The lines represent the regions which include 90\% of the scan points for the $A \rightarrow \tau^+ \tau^-$ and $B_s \rightarrow \mu^+ \mu^-$ decays at the LHC and the dark matter direct detection at the XENON experiment.}
  \label{fig:pmssm3}
\end{figure}

In Fig.~\ref{fig:pmssm3} the points fulfilling the Higgs constraints are displayed in the plane ($M_A ,\tan \beta$). We observe that imposing the value of $M_h$ selects a broad wedge at relatively large $M_A$ and moderate to large values of $\tan\beta$, extending beyond the projected sensitivity of the searches for the $A^0 \to \tau^+ \tau^-$ decay but also that of direct DM detection and would be compatible with an SM--like value for the rate of the $B_s^0 \to \mu^+ \mu^-$ decay. 
Imposing in addition that the yields in the $\gamma \gamma$, $W^+W^-$ and $Z^0Z^0$ final states satisfy the conditions 1$\le R_{\gamma \gamma} <$3 and 0.3$< R_{W^+W^-/Z^0Z^0} <$2.5, the wedge in the ($M_A$, $\tan \beta$) plane is further restricted, as can be seen from the right hand side of Fig.~\ref{fig:pmssm3}.

\begin{figure}[!t]
\centering
\includegraphics[width=7.5cm]{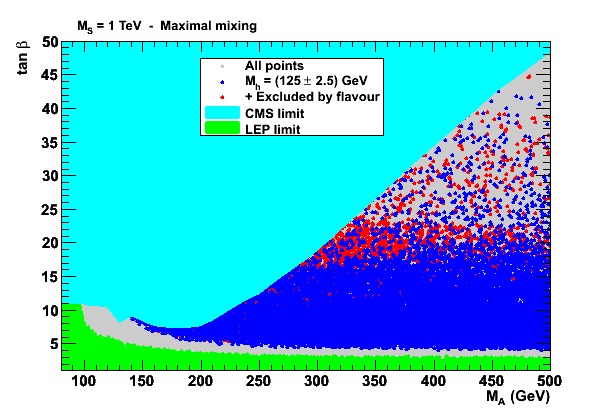}
\includegraphics[width=7.5cm]{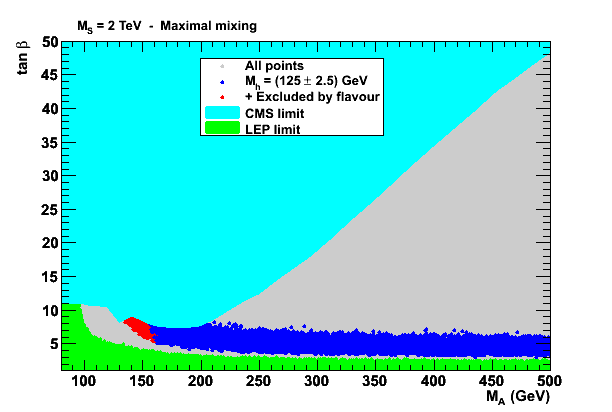}
\includegraphics[width=7.5cm]{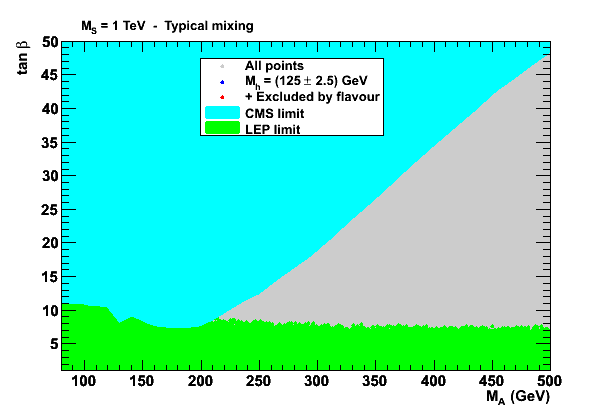}
\includegraphics[width=7.5cm]{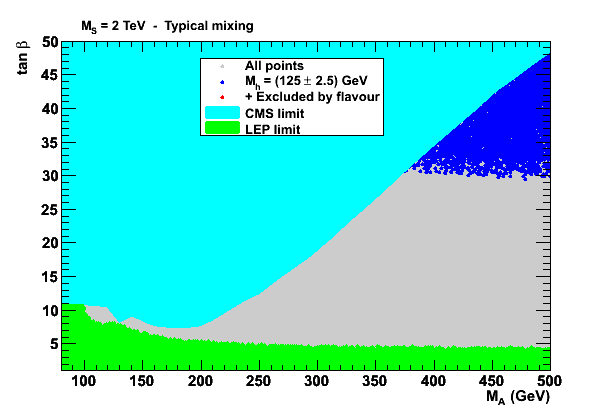}
\includegraphics[width=7.5cm]{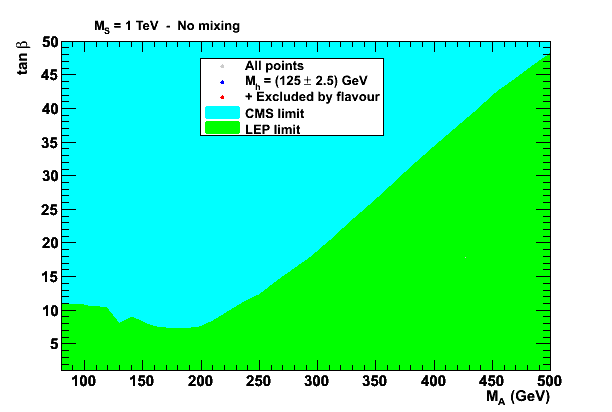}
\includegraphics[width=7.5cm]{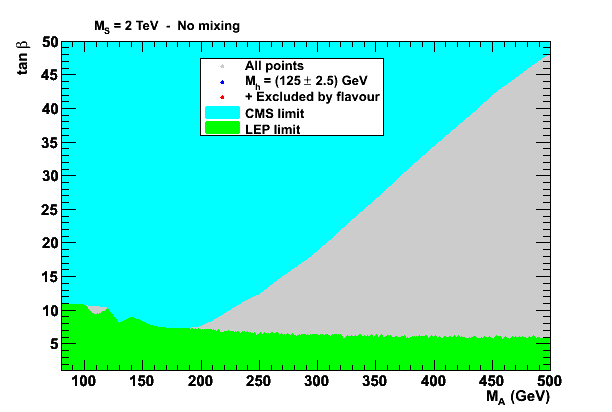}
\caption{Points in the ($M_A$, $\tan \beta$) plane for the maximal mixing (top), typical mixing (center) and no mixing (bottom) scenarios for $M_S=1$ TeV (left) and 2 TeV (right). The ``LEP limit'' zone correspond to exclusion by Higgs searches at LEP, whereas the ``CMS limit'' zone is excluded by the CMS searches for $A^0 \to \tau^+ \tau^-$ with 4.6~fb$^{-1}$ of data.}
  \label{fig:pmssm4}
\end{figure}

Finally we highlight the consequences of the $h^0$ mass limit for three benchmark scenarios, namely the maximal, typical and no-mixing scenarios. The results in the plane ($M_A$, $\tan \beta$) are presented in Fig.~\ref{fig:pmssm4}, where for comparison we show also the LEP Higgs search limits \cite{Schael:2006cr} and the CMS limits from $A^0 \to \tau^+ \tau^-$ searches with 4.6~fb$^{-1}$ of data \cite{Chatrchyan:2012vp}. In addition, we apply flavour physics constraints from $B_s \to \mu^+ \mu^-$, $B \to \tau \nu$ and $b \to s \gamma$. As expected, the no-mixing scenario is excluded for $M_S =$ 1 TeV by the combination of LEP and CMS limits, and there is no solution for $M_h$ in the accepted interval for $M_S =$ 2 TeV. The situation is similar for the typical mixing scenario for  $M_S =$ 1 TeV where no solution is found, whereas for $M_S =$ 2 TeV a small corner survives at large $\tan\beta$ and large $M_A$. On the other hand, as we have shown earlier, the maximal mixing scenario provides more solutions especially for $M_S =$ 1 TeV. For larger $M_S$ only a narrow band remains at small $\tan\beta$ (around 5). Flavour physics constraints further exclude this region where $M_A \lesssim$ 160 GeV. The remaining allowed region in the ($M_A$, $\tan \beta$) plane is therefore strongly narrowed down when one takes into account the constraints from $M_h$. 
%
\section{Conclusions}
The Higgs boson searches at the LHC, in conjunction with the flavour physics limits place highly
constraining bounds on the MSSM parameters. The impact is very strong in particular in the constrained MSSM scenarios and several scenarios such as mAMSB and mGMSB are disfavoured as they lead to a too light $h^0$ (for $M_S <$ 3 TeV). In the pMSSM on the other hand, strong restrictions can be set on the mixing in the top sector. In the ($M_A$, $\tan \beta$) plane we observe a favoured wedge corresponding to rather large values of the $A^0$ mass and moderate to large values of $\tan\beta$. Further imposing the yields in the Higgs decay rates, the wedge becomes more pronounced and the fraction of accepted points gets reduced preferentially at large sparticle masses. Focussing more on specific benchmark scenarios we see that the no-mixing and typical mixing scenarios are severely restricted, while the maximal mixing scenario provides more solutions, and imposes competitive constraints with the CP-odd and charged Higgs searches on $M_A$ and $\tan \beta$.   
%
\section*{Acknowledgements}
I would like to thank the organisers of the Moriond Electroweak meeting 2012 for their invitation, and A. Arbey, M. Battaglia and A. Djouadi for their collaboration.
%

\end{document}